# My First Five Years of Faculty Career at the University of Delaware

Xiang-Gen Xia

University of Delaware, Newark, DE 19716, USA

**Abstract**

In this short article, I would like to briefly summarize my research in the first 5 years in my university academia life in USA. I think that my research results obtained in these 5 years are the best in my career, at least which I like the most by myself. I wish that my experience in my junior academia career could be of some help to young researchers.

**1. Joining University of Delaware**

This year marks my 30th year at the University of Delaware. Looking back, I have many recollections and thoughts. Let me first recollect my first five years of research at the University of Delaware.

In 1983, I graduated from the Department of Mathematics at Nanjing Normal University, Nanjing, China, and was admitted to the Master's program in Mathematics at Nankai University, Tianjin, China. After graduating with a Master's degree in Mathematics from Nankai University in 1986, I remained on the faculty there. In 1988, I started to pursue my Ph.D. in Mathematics at the University of Cincinnati, USA. Since I did my first research in signal processing before I came to USA, I had been always interested in signal processing in electrical engineering. As such, in 1990, I transferred to the Department of Electrical Engineering at the University of Southern California, USA, for my Ph.D. While pursuing my Ph.D., my dream was to become a professor at a leading American research university. However, when I graduated with my PhD in 1992, there were very few university faculty positions in the United States. After nearly three years of postdoc work and industry, I finally received an offer in late 1995 for a tenure-track (TT) assistant professorship in the Department of Electrical Engineering at the University of Delaware (the most junior, regular TT professorship at a university). I remember receiving the offer letter on December 24, 1995 (I had just returned from Nanjing after attending an international signal processing conference hosted by Professor He Zhenya in Southeast University,

Nanjing, China). The offer required a response by January 2, 1996. I didn't have much time to think about it, and I accepted it right away without hesitation.

I remember that at the time, I had 47 published and accepted journal papers. Note that by 1995, I had almost 10 years of academic paper publication history. Here I'm not referring to counting the number of papers.    Refereed academic journals were still quite respectable, rigorous, and responsible back then. I believe each of my papers had some novel results. It was precisely because of this foundation that I was able to achieve the next five years research results. My University of Delaware job interview presentation in 1995 was on the latest results in filterbank precoded intersymbol interference channels. In 1995, despite the booming communications industry, university academia open positions were still sparse, which is because that in terms of career prospects, academia typically lags behind industry. IEEE Spectrum advertised very few vacancies, and the University of Delaware was one of the few. Starting from 1996, university faculty positions suddenly increased significantly, and the dot COM craze had officially arrived, making finding a position in academia much easier.

## 2. Research After Joining University of Delaware

After joining the University of Delaware in the fall of 1996, I continued to work six days a week (Sunday through Friday), at least 12 hours a day. It wasn't because of any pressure. I was simply passionate about research, interested in producing new results, and had many good research problems to work. Honestly, I never worried about getting tenure. Perhaps it was my good fortune to receive funding in my first year at the university. Unlike in industry, a university professor had a lot of freedom in research. There were no many time-consuming meetings, and the university was very supportive of young faculty members, not requiring them to do too much services. I spent my days commuting between home and the university office, one line with two points, completely dedicated. I never ask students to work a certain amount of time each day, or to come to the office all the time. It's important to understand that research in our field is intellectually demanding, and you can't force it. But every semester, I emailed them about my work schedule, as mentioned above.

The book and papers listed below were all written by myself after joining the University of Delaware, deriving formulas, performing simulations, and writing. They were all published within the first five years of my tenure at the University of Delaware. They do not include any papers where I'm not the first author (e.g., papers written by my students). I think that there are two types of new scientific research. One type requires reading numerous books and papers until you're at the forefront of a discipline before you can begin working on it. Without reading so many books and papers, you won't recognize many of the notations and won't understand what others are talking about. This is common in mathematical and physics disciplines, such as differential geometry, algebraic geometry, and algebraic topology. The other type involves creating genuine novelty and innovation within commonly well-known basic topics, which is more likely to be seen in engineering. I think that I have at least four results of the latter type, all of which I started working on myself after joining the University of Delaware, and all of which were published in journals within the first five years.

**The first** is a new pulse [2] (called Xia pulses/filters now), which satisfies the Nyquist property whether with or without matched filtering. The raised cosine pulse (after square root) that everyone is familiar with in college textbooks is Nyquist only after matched filtering. These two pulses are closely related.

**The second** is Vector OFDM (VOFDM) [20]. As we all know, OFDM was developed in the 1960s. In fact, as I mentioned in one of my written articles before ("Recall Gabor Communication Theory and Joint Time-Frequency Analysis," https://arxiv.org/pdf/2509.02724), the mathematical form of OFDM can be traced back to Gabor's paper in 1946. It is now ubiquitous and all WiFi and mobile phone communications use it. VOFDM is the most general extension of OFDM to inter-symbol interference (ISI) channels, spanning the single carrier and multi-carrier, time domain and frequency domain possibilities, depending on the different requirements for demodulation complexity. Seventeen years after the first publication of VOFDM in the 2000 ICC conference proceedings, OTFS emerged, with its transmission being identical to that of VOFDM.

**The third** is the generalized and robust Chinese remainder theorems [12, 14]. Only the generalized Chinese remainder theorem is listed here, and the robust Chinese remainder theorem was not published until later. This was the beginning of my research on Chinese remainder theorem, which I am still working on. The Chinese remainder theorem is well-known and ancient, and is the most famous and important mathematical result from China in the world. It has important applications in many fields.

**The fourth** is the discrete chirp-Fourier transform (DCFT) [16]. Perhaps all college students are familiar with the discrete Fourier transform (DFT). DFT is also an essential tool for OFDM and is extremely well-known. DFT is used to match sine and cosine waves, that is, to match constant frequency signals. However, if we slightly generalize the signals from constant-frequency signals to linear frequency-modulated

(LFM) signals, also known as linear chirps, the DFT immediately breaks down. The DCFT is specifically designed to match LFM signals. Interestingly, it performs optimally when the discrete signal length is a prime number. The DCFT is actually related to the fractional Fourier transform (FrFT), both of which can be used to match and only match LFM signals, although the original motivation for FrFT is not to match LFM signals. However, the DCFT tells us when it is optimal.

Before joining the University of Delaware, I also conducted some research on FrFT. For example, I was the first to study the sampling theorem for band-limited signals in the fractional domain (published in *IEEE Signal Processing Letters* in the March of 1996), and I proved that a signal cannot be simultaneously band-limited in the fractional domain at two different rotation angles. This theorem was proved by using one of the Paley-Wiener theorems as I mentioned in one article of mine on Paley in Chinese a few days ago. Later, when I was working at the Hughes Research Labs., I used FrFT to propose generalized marginal time-frequency distributions, generalizing the marginal distributions in the time and frequency domains to the marginal distributions in any two fractional domains, and obtained a necessary and sufficient condition for them. This result was published in *IEEE Transactions on Signal Processing* in the November of 1996.

Another result that I also consider fundamental is the analytical analysis of the signal-to-noise ratio (SNR) in the joint time-frequency domain [4, 11]. As we all know, for many signals, after a time-frequency transform, such as the short-time Fourier transform (STFT), they are not visible in the time or frequency domain, but are visible in the time-frequency domain. In other words, their SNR in the time-frequency domain increases. So how do we characterize this property analytically? In other words, how do we calculate the SNR in the time-frequency domain? To calculate the SNR, we must first clarify the reasonable definition of the SNR. The SNR we usually use is actually defined for stationary signals, and is always calculated in an average sense, that is, the average power ratio of the signal to the noise. However, when we see the signal in the time-frequency domain, it is no longer stationary. To this end, I first defined the 3dB SNR, then calculated the SNR in the STFT and pseudo Wigner-Ville domains, and derived a formula showing that the increase in SNR in the joint time-frequency domain is proportional to the sampling rate of the signal in the time domain.

One more result, although may not be considered fundamental, is theoretically quite interesting. It is an application of filterbanks to blind (or semi-blind) equalization of ISI channels [13, 15, 17], namely polynomial ambiguity resistant precoders (PARP), *P(z)*:

$$Y(z)=H(z)P(z)X(z)+W(z),$$

where *H(z)* is an unknown channel, *X(z)* is an information signal wanted to be sent by the transmitter, *P(z)* is a precoder at the transmitter, *Y(z)* is the signal received by the receiver, and *W(z)* is the added noise. If *P(z)* is a PARP, then the receiver can use *Y(z)* and *P(z)* to determine *X(z)* uniquely in the sense that the difference is only a common constant. PARP are about polynomial matrices where all components are polynomials and the concept of PARP is new in mathematics as well. We have characterized and constructed such PARP.

The papers listed below cover a number of other topics, such as those related to radar, wavelets, delay Doppler channels, and Gabor transforms, which I will not introduce in this article. Here again I am not trying to count the number of papers and I believe that each of the papers I list contains truly new results. I joined the University of Delaware in the fall of 1996 and was fortunate to receive tenure in 1999. Therefore, the last two of the first five years I described above were my post-tenure years. In fact, tenured-or-not did not make much difference to me anymore, since I had already become a life-time dedicated researcher since a very long time ago!

Also, it was during these five years that my second child was born.

## 3. Conclusion

Although this year marks my 30th year at the University of Delaware, I feel most satisfied with the research results from my first five years there, and I always enjoy rereading them. I could imagine that rereading my old papers will be a real pleasure after my retirement. Finally, I sincerely hope that my recollections above will be of some help to young professors.

**Refereed Journal Publications**

[1] X.-G. Xia, "System identification using chirp signals and time-variant filters in the joint time-frequency domain," *IEEE Trans. on Signal Processing*, vol. 45, no. 8, pp. 2072-2084, Aug., 1997.

[2] X.-G. Xia, "A family of pulse shaping filters with ISI-free matched and unmatched filter properties," *IEEE Trans. on Communications*, vol. 45, no. 10, pp. 1157-1158, Oct. 1997.

**Refereed Conference Proceeding Publications Not Covered in Any Papers Above**

**Books**